\documentstyle[aaspp4,11pt,psfig]{article}

\def\dw#1{}			
\def\jm#1{}

\begin{document}


\title{Small Scale Structure at High Redshift: \\
\  \\
I. Glimpses of the Interstellar Medium at Redshift $\sim 3.5$ 
\altaffilmark{1}}

\today
\vskip 1.5cm

\author{Michael Rauch\altaffilmark{2,3,4}, Wallace L.W. Sargent\altaffilmark{2},
Tom A. Barlow\altaffilmark{2}}
\altaffiltext{1}{The observations were made at the W.M. Keck Observatory
which is operated as a scientific partnership between the California
Institute of Technology and the University of California; it was made
possible by the generous support of the W.M. Keck Foundation.}
\altaffiltext{2}{Astronomy Department, California Institute of Technology,
Pasadena, CA 91125, USA}
\altaffiltext{3}{Hubble Fellow}
\altaffiltext{4}{ New address: European Southern Observatory, Karl-Schwarzschild-Str. 2,
85748 Garching, Germany}

\medskip 
\vskip 1.5cm

{Subject Headings:  galaxies: kinematics and dynamics ---
galaxies: structure --- interstellar medium --- quasars: absorption lines} 

\vskip 2.0cm
\centerline{\it accepted for publication by the Astrophysical Journal}


\vfill

\pagebreak

\begin{abstract}
We have obtained high resolution (FWHM = 4.4 kms$^{-1}$) Keck HIRES spectra of
images A and C of the gravitationally lensed QSO Q1422+231 (z$_{em}$ = 3.628).
The images are separated by 1.3" on the sky. In an absorption system
at z$_{abs}$ = 3.538 gas column density variations by an
order of magnitude and velocity shear on the order of 10 kms$^{-1}$
are observed in 
the low ionization (SiII, CII) lines.  The transverse separation in 
the absorbing cloud is estimated to be as small as 26h$_{50}^{-1}$ parsec,
corresponding to an effective angular resolution of 4 milliarcseconds 
as seen from the Earth.
In contrast, the high ionization (CIV) gas appears mostly featureless
and thus must be considerably more extended.  The abundances of the
elements carbon, silicon and oxygen appear to be close to the solar
values. 
The observation provides the first spatially and kinematically resolved
probe of the interstellar medium at high redshift on scales small
enough to be influenced by individual stars or star clusters. 
The
mass associated with the low ionization "cloudlets" is likely to be
less than about 3000 $M_{\odot}$ and possibly less than 1 M$_\odot$.
The
velocity shear seen across the lines of sight is too large to be caused
by galactic bulk motion, so the velocity field of the low ionization
gas must be strongly influenced by small scale local gasdynamics. While
it cannot presently be excluded that the disturbances of the gas are due to high
velocity outflows from the background QSO, the observed velocity and
density structure of the z=3.538 system is consistent with our line
of sight running through an expanding shell of gas, possibly a
supernova bubble or a stellar wind.

\end{abstract}

\pagebreak

\section{Introduction}

Observations of intervening absorption in multiple, gravitational
lensed QSO images (e.g., Young et al., 1981; Weymann \& Foltz 1983; Foltz et
al.  1984; Smette et al.  1993,1995; Bechtold \& Yee 1995; Zuo et al.
1997; Michalitsianos et al. 1997, Rauch 1998, Petry et al. 1998, Lopez et al 1998)
provide a very useful tool for investigating the distribution of matter
at high redshift.  For example, the typical sizes of various types of
high redshift cosmic structures can be measured in a statistical sense
from the numbers of coincident and anticoincident absorption systems in
both lines-of-sight (LOS); depending on the amount of spectroscopic detail
available, additional inferences can be drawn about the baryonic
density and velocity fields and their spatial variations. Lensing by an
intervening galaxy enlarges - like a magnifying glass - any structures
on spatial scales from virtually zero extent (down to the size of the
QSO continuum emission region) up to the maximum beam separation (tens
of kpcs, at the redshift position of the lensing galaxy) by opening the
projected angle between the separate QSO images to a few arcsecs, where
they can be spatially resolved with ground-based telescopes.

Unfortunately, gravitationally lensed images of a given object tend to have
very different brightnesses among each other. Accordingly, the observation of
at least two images sufficiently bright to be observed while simultaneously resolving
the absorption lines had to wait for the light-gathering power of the
new generation of large optical telescopes (Rauch 1998; Rauch, Sargent
and Barlow, in prep.).

In the first of a series of papers on the small scale properties of high
redshift (z$\sim 3$) structures we report here on
spectra obtained with the HIRES instrument (Vogt et
al.  1994) at the Keck I telescope. The A (V=16.7) and C (V=17.3)
images of the QSO 1422+231 ($z_{em}$=3.628, Patnaik et al. 1992) have
been observed during three runs in the spring of 1998, with total exposure
times of 19600 (33200) seconds for the A image (C image).  As the
angular separation between the images is only about 1.3 arcseconds,
only spectra obtained when the seeing was less than $\sim$ 0.6" were
included, a condition which was satisfied at the end of a few nights
when the seeing sometimes dropped below 0.5".  A
$7\times0.574$ arcsecond wide decker (giving a spectral resolution
FWHM=4.4 kms$^{-1}$) was placed separately on each image, with the
position angle set such as to minimize contamination from the other
images.  When the seeing threatened to deteriorate the slit was shifted
slightly off the center of an image away from
the other images.  We estimate the contamination of our exposures of
the A and C images by each other to be at most a few percent. The
absence of significant contamination can also be judged  from the fact
that strong differences are observed between the LOS in the appearance of some absorption systems.

The redshift of the lens is believed to be z$_l$=0.338 (see Kundic et
al 1997, Tonry 1997). Combined with the small angular splitting, this
low redshift leads to transverse beam separations of less than 100 pc
as far as 10 Mpc or more away from the QSO where any intervening matter
is less likely to be dynamically influenced by the quasar.

\section{The z=3.538  Metal Absorption System}

\subsection{Description}

Most of the absorption lines in the spectra of the A and C images look
very similar because of their small separation (see also Bechtold \&
Yee 1995, Petry et al.  1998, Rauch 1998). However, striking
differences are found in the appearances of some low ionization metal
absorption systems. In particular, a strong system at z=3.538 for which
both low (OI, SiII, CII) and high (CIV, SiIV) ionization lines can be
observed appears to hold clues as to the small scale properties of the
absorbing gas clouds (Fig. 1).  The origin of the (proper) velocity
scale  in the figure has been arbitrarily set to coincide with redshift
z=3.53792.  The beam separation here is only $\Delta r \approx$
26h$_{50}^{-1}$ pc, corresponding to an effective angular resolution of
3.8 milliarcsecs (for $q_0$=0.5).  The system is blueshifted with
respect to the QSO's (CIV) emission line by about 6000 kms$^{-1}$ in
the rest frame of the QSO.  The actual velocity difference is likely to
be even somewhat larger because of the blueshift commonly observed
between broad and narrow ("systemic") emission lines (Espey 1993, and
references therein).  It is not {\it a priori} clear whether the
absorbing gas clouds are being ejected from the QSO (in which case all
or a part of the redshift difference would be due to the velocity and
the gas could be very close to the QSO), or whether they belong to a
separate galaxy, a question we will discuss below.  If independent from
the QSO,  the absorbing clouds are at a proper luminosity distance of
11.6 h$_{50}^{-1}$ Mpc from the QSO, if we adopt z$_{em}$ = 3.628 as
the emission redshift, consistent with an origin in an intervening
galaxy unrelated to the QSO.

\begin{figure}[tbp]
\centerline{
\psfig{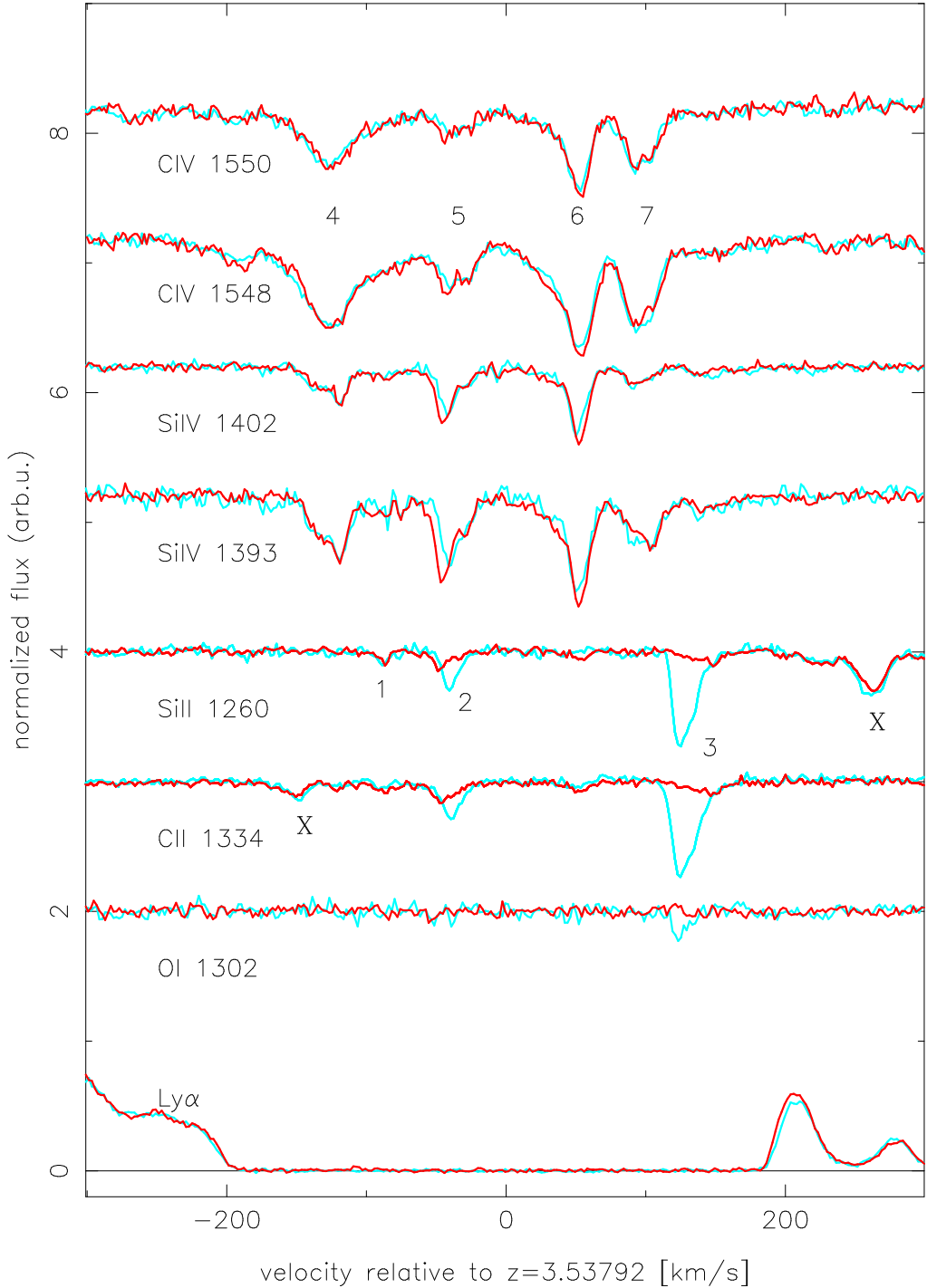}}   
\caption{\small Absorption system at z=3.538 as seen in the A (grey line)
and C (black line) images of QSO 1422+231. The plot shows the continuum normalized flux in various ions
versus a relative velocity along the LOS (arbitrary zeropoint), for 8 transitions. The sections are offset in the y-coordinate by suitable amounts for
clarity. For convenience, some components discussed in the text are
singled out by numbers. The rightmost absorption line ('$X$' near 260
kms$^{-1}$) in the SiII 1260 region is an accidental CIV 1548 interloper
(at z$\sim$ 2.7), unrelated to the system. the '$X$' component in the CII section
of the spectrum is a similar, weak stray CIV at z=2.91 (A. Boksenberg, priv. comm). \label{allions}}
\end{figure}

Figure 1. shows the absorption complex as it appears (going from top to
bottom) in the CIV doublet, SiIV doublet, SiII, CII, OI and
Ly$\alpha$.  In each case, the solid line is the spectrum of the C
image, the grey line that of the A image. For further reference we have
numbered the strongest low ionization (1 -- 3) and high ionization
(4 -- 7) components.

For this particular system the high and low ionization lines exhibit
very different component structures. While the total velocity extent is
similar ($\sim 300$ kms$^{-1}$) in both cases, only  component 2
appears to be clearly present in all ions.  It is further interesting
that the CIV components in the two images seem to be identical within
the errors. This lack of structure indicates that the typical gas cloud
giving rise to CIV lines is larger than a few tens of parsecs. From
more widely separated lines of sight (Rauch 1998; Rauch, Sargent and
Barlow 1998) we also know that the structures producing individual CIV
lines differ in column density by 50 \% or more over a few hundred
parsecs to a kiloparsec, so this range appears to be characteristic of
the typical sizes of CIV ``cloudlets''.  The SiIV absorption patterns
are also very similar between the two LOS, except for some slight but
significant differences in the column density and/or velocity structure
of component 2. The observed differences can be interpreted as a shift
in column density weighted velocity by $\Delta v$ = $v_A - v_C \approx
3$ kms$^{-1}$ between the two lines of sight. A similar shift is also
visible in SiII 1260 and CII 1334 ($\Delta v \approx 5$ kms$^{-1}$ for
component 2, $\Delta v \approx -16$ kms$^{-1}$ for component 3.
However, the most striking difference is found between the low
ionization column densities across the LOS (see table 1).  Component 2
has a SiII (CII) column 2.1 (2.4) times larger in the A image; the
strongest low ionization component 3 is even 10.5 (10.0) times stronger
in the A than in the C image.

\subsection{Physical State and Mass\altaffilmark{1}}

\altaffiltext{1}{Unless other components are  mentioned explicitly the analysis below
will be focussed on the strongest low ionization component (\# 3) in
the A image.}

Constraints on the ionization parameter, metallicity and density of the
absorbing gas can be obtained by using observations of several ions
(e.g., Bergeron \& Stasinska 1986, Chaffee et al. 1986, Steidel 1990,
Donahue \& Shull 1991).  First, we note that the system \#3 is
apparently optically thin to ionizing radiation, which is somewhat
surprising given the strength of the low ionization lines SiII and CII.
From a direct Voigt profile fit to several lines of the Lyman series
(up to Ly 7) we get $\log N$(HI)=16.05$\pm$0.15 (with $N$ measured in units
of cm$^{-2}$) for component \# 3,
in LOS A; the error gives only the uncertainty of the fit. For most of the Lyman
series except Ly$\alpha$ there may be additional systematic uncertainties in both
the continuum and zero level placement, and the systematic error may be
considerably larger than that quoted above.  
Unfortunately the spectra are quite noisy in the blue and they do not cover the Lyman
limit of the system. 
Thus the HI column density
is not very precisely known, although trials (assuming the same Doppler
parameter $b$ = 21 kms$^{-1}$) show that a column density of $\log N$=16.5
is clearly too high. Accordingly, unless the Doppler
parameter is much smaller and the line is unresolved even at our
resolution, the aforementioned value is a conservative upper limit.
(The system in question has been detected in earlier observations of
Q1422+231 by ourselves and independently by Songaila \& Cowie (1996)
who find a total  HI column density $N$(HI)=2.3$\times 10^{16}$
cm$^{-2}$.  However, these older spectra were not taken for the purpose
of splitting the contributions from separate images.) A comparison of
the absorption line strength between low and high ionization species
shows immediately that the gas of component 3 must be of rather low
ionization.  We have used the results of the photoionization
calculations of Donahue \& Shull (1991; hereinafter DS) summarized in
their figure 7 (they used a typical QSO spectrum after Matthews \&
Ferland (1987), and a metallicity $Z$ = 0.1 $Z_{\odot}$, assuming solar
relative abundances). Our observed
column density ratios of neutral oxygen to singly ionized carbon,
\begin{eqnarray}
\log \frac{\mathrm N(OI)}{\mathrm N(CII)}  = -0.45 \pm 0.06
\end{eqnarray}
imply an ionization parameter
\begin{eqnarray}
\log U = \frac{n_{\gamma}}{n} \simeq - 4.4,
\end{eqnarray}
in agreement with the upper limit from the SiII/SiIV ratio 
\begin{eqnarray}
\log U \leq - 3.4.
\end{eqnarray}
The total H density $n$, again according to DS, is then
\begin{eqnarray}
n\simeq 1.6 \left(\frac{J_{912}}{10^{-21}}\right){\mathrm cm}^{-3}, \ \ \ {\mathrm and\ \ \ } n \geq 0.16 \left(\frac{ J_{912}}{10^{-21}}\right){\mathrm cm}^{-3},
\end{eqnarray}
respectively, for the two $U$ values where $J_{912}$ is the intensity of the radiation field at the Lyman
limit
{\em averaged} over $4\pi$ steradians,in units of ergs cm$^{-2}$ Hz$^{-1}$ s$^{-1}$ sr$^{-1}$.

In spite of its low 
HI column density the neutral fraction of the gas must be as large as 0.15 (0.014) for $\log U$ = --4.4 (--3.4). 
For $J$ = 10$^{-21}$ ergs cm$^{-2}$ s$^{-1}$ Hz$^{-1}$ sr$^{-1}$,
and using the inferred H density, HI column density, and neutral fraction 
we get a size estimate along the line of sight,
\begin{eqnarray}
L \sim 0.015\ {\mathrm pc}, \ \ \ {\mathrm and} \ \ \ L \leq 1.6\ {\mathrm pc}.
\end{eqnarray}
It is important to bear in mind that these derivations are assuming  
thermal photoionization equilibrium with a QSO radiation field, 
which may not be very realistic.

An independent upper
limit on the density can be obtained from the absence of the (J=$1/2 \rightarrow 3/2$) fine structure
transitions in CII*($\lambda 1335.7 \AA $) and SiII*($\lambda 1264.7 \AA $)
(Bahcall \& Wolf 1968; York \& Kinahan 1979).
Here CII* gives a somewhat tighter upper limit than SiII*.
We observe 
\begin{eqnarray}
\frac{N({\mathrm CII}^*1335.7)}{N({\mathrm CII\ }1334.5)} \leq 0.04,
\end{eqnarray}
from which we get an approximate upper limit on the electron density
of 
\begin{eqnarray}
n_e \leq 1.5 \times \left(\frac{T}{10^4 K}\right)^{0.5} {\mathrm cm}^{-3}
\end{eqnarray}
(valid for temperatures above a few hundred Kelvin).
Assuming that most of the gas is ionized, $n \simeq n_e$, an upper limit close
to the maximum density allowed by the DS photoionization models is obtained,
implying that the density must be somewhere in the range
\begin{eqnarray}
0.16 \leq n \leq 1.6\ {\mathrm cm}^{-3}
\end{eqnarray}

We know that the transverse size of the clouds (i.e., the distance over which
the column density decreases by a factor 10) is on the order of 26 pc.
The gas mass associated with the cloud can then be estimated very crudely 
to lie between that of a homogeneous, cylindrical slab of size L $\times \pi (\Delta r)^2$
(assuming the smallest value for L and a radius $\Delta r$) on one hand, 
and (ignoring the photoionization size estimates) a spherical cloud with a radius
$\Delta r$ and the high density limit on the other: 
\begin{eqnarray}
0. 4\leq M \leq 2700\  M_{\odot}.
\end{eqnarray}
The assumption of homogeneity is of course questionable. In any
case, the scale indicates that, for the first time we are resolving
structure on scales relevant to individual star clusters or even single stars in a
high redshift galaxy.

The remaining absorption components visible in CII and SiII appear to have less extreme
ionization states. Component 2  appears in all ions except OI.
The ionization parameter (in LOS A) is measured to be
\begin{eqnarray}
\log U  \simeq - 2.7 \ \ \  {\mathrm(from\ \  SiII/SiIV)},\ {\mathrm and\ } 
\log U \simeq - 2.8 \ \ \ {\mathrm(from\ \  CII/CIV)},
\end{eqnarray}
typical for stronger high ionization systems when both SiIV and CIV are detected.
In deriving U it is assumed that the singly ionized and triply ionized gas clouds
have the same spatial extent. In general, this is not
true as can be seen directly from the larger sizes inferred for most CIV systems.
However, for this particular component the CIV (and SiIV) both seem to
trace the shift in the velocity between the LOS seen in the low ions
reasonably well, so most of the CIV here may indeed occupy a similar
volume as the CII.

The other high ionization components 4, 6 and 7 are unexceptional in
that there are hardly any column density or velocity differences
between the LOS, and the ionization level is consistent with the
observed range of CIV systems. For example, for component 6, $\log U
\simeq$ --2.5.  This is in spite of the relative proximity of the QSO; one would expect a
higher degree of ionization if the
QSO's radiation field dominates over the metagalactic background.  
We have tried to detect NV absorption, a characteristic sign of 
gas associated with the large ionizing intensity near the QSO,  at the positions of
the known CIV components, weighted by the model CIV column density ratios.
Only an upper limit of $\log N$(NV) $<$ 12.25 (1-$\sigma$) for the {\em total}
NV column of the z=3.538 system was found. Compared with the total CIV column
$\log N$(CIV) = 14.34$\pm 0.03$ this leads to an upper limit on the averaged
ionization parameter $\log U <$ --3.2. This is somewhat less than the U values measured
for the individual high ionization components from other ion ratios (see above)
and it may imply that either the nitrogen abundance is lower than solar relative
to C or Si, or that there is a lack of high energy photons capable of ionizing NIV,
as is indeed expected for a radiation field dominated by starlight.

\subsection{Temperatures}

In the case of the high ionization component 4, the absorption line can
be modelled as three relatively broad components.  Even the apparently
narrow "Gaussian core" of the line requires a component with b = 14
kms$^{-1}$, corresponding to a temperature of $>$ 1.5$\times$10$^5$ K,
if thermal.  This b value is considerably larger than the typical
thermal width of CIV lines ($b_{median} \sim 7 $kms$^{-1}$,
corresponding to $T \sim 4\times10^4$ K; Rauch et al.  1996).  The broad wings
are not repeated in the SiIV profile, perhaps implying that additional
hot (collisionally ionized) gas contributes to the CIV profile.
However, cooler gas is present as well. Lines 6 and 7, while
also containing assymmetric broad wings, do have cores more similar to
typical CIV widths : line 6 has a dominant component with $b$ = 8.9
kms$^{-1}$, line 7 one with $b$ = 6.4 kms$^{-1}$; both lines are
consistent with mostly thermal broadening caused by photoionization.

The low ions clearly show lower temperatures. Returning to component 3,
the line appears to be a blend of several components, which can be
modelled as three Voigt profiles. The sharp blue wing of CII (SiII) is
well-fitted by $b_{\mathrm CII}$ = 4.10$\pm 0.50$ kms$^{-1}$
($b_{\mathrm SiII}=$3.95$\pm 0.33$ kms$^{-1}$). The similarity of the
values indicates that bulk motion is more important than thermal
broadening.  Thus we have an upper limit to the temperature, $T < $
1.2$\times$10$^{4}$ K.

\subsection{Metallicities}

The metal abundance of at least one component of
the system appears to be rather high. Using the measured OI/CII ratio of component
3 together with the relations between ion abundance and U parameter
from DS, we predict $\log$($N$(OI)/$N$(HI)) $\simeq$ --4.0, whereas, from the
measured N(HI) we would expect $\log$($N$(OI)/$N$(HI)) $\simeq$ --2.8. Similarly,
the predicted $\log$($N$(SiII)/$N$(HI)) $\leq$ --4.2, in contrast to the measured
--3.2.  The DS model was for a metallicity $Z=0.1 Z_{\odot}$, so
raising the abundances to approximately solar gives consistent values.
The relative abundances appear to be close to solar as well.
Assuming that the ionization parameter $\log U$ = --4.4 (derived from
the OI/CII ratio) holds, the CII/SiII ratio is expected to be $\log$
($N$(CII)/$N$(SiII)) $\simeq$ 1.1, not very different from any of the
observed values for $\log$ ($N$(CII)/$N$(SiII)) given in table 1. Even if if the
O/C would be higher than solar (as in metal poor stars) 
so that the true ionization
parameter would be higher than estimated here, the conclusion that Si/C is solar would
still hold, as  CII/SiII is rather insensitive to the ionization
parameter.

\section{Possible Interpretations: Ejected Gas or Intervening Galaxy ?}

Given the relative proximity of the system to the QSO one may wonder
whether the gas is possibly ejected from the QSO host galaxy, and therefore
to be classified among the so-called $z_{\mathrm abs} \simeq z_{\mathrm
em}$ systems.  The main properties of these systems
are: absorption redshift close to or beyond the emission redshift of
the QSO; only partial coverage of the QSO's emission region; high
(super-solar) metallicity; and a high ionization state caused by the AGN
radiation field, evident for example through the presence of NV.
Petitjean, Rauch \& Carswell (1994) have found high metallicity systems
out to 13000 kms$^{-1}$ from the QSO emission redshift, and with a
velocity separation of $\sim 6000$ kms$^{-1}$ the z=3.538 system is within
that range. However, the HI lines corresponding to components 2 and 3
do go down to zero intensity so the clouds must be covering the  QSO 
continuum and line-emitting regions completely.
Moreover, we do not find NV in any of the components down to limits
lower than any of the individual NV upper limits or detections in the
Petitjean et al data, nor do we detect (see above) high [Si/C] ratios
as did Savaglio, D'Odorico \& M\o ller (1994) in their $z_{\mathrm abs}
\simeq z_{\mathrm em}$ sample.  Models of galactic chemical evolution
(Matteucci \& Padovani 1993) show that the presently observed abundance
pattern (C, Si solar, N $\leq $solar), unlike the pattern seen in many
QSO emission line spectra and many $z_{\mathrm abs} \simeq z_{\mathrm
em}$ systems (N, Si enhanced by up to an order of magnitude, C and O
solar) can be produced in the very early phases of galactic
nucleosynthesis, shortly  after the onset of the first star burst in an
elliptical galaxy ($t < $ a few $\times$ 10$^8$ years), even before the
QSO abundance anomalies (e.g., Hamann 1997) are established.  Nevertheless, it cannot be
excluded that the system is ejected or at least dynamically influenced
by the QSO. Radiation pressure or ram pressure from outflows may
explain why the lower density, high ionization gas seems to be
blueshifted with respect to the denser low ionization clumps, from
which the CIV gas may have been ablated like a cometary tail.

If, however, the absorption system is not influenced dynamically by the
QSO, where do the small scale velocity variations across the LOS come
from ?  On a scale of 26 pc, the velocity shear observed most
prominently in components 2 and 3 (about 10 km/s) is on the order of
${\Delta v_{||}/\Delta r}\sim 400$ kms$^{-1}$ kpc$^{-1}$ (across the sky, and projected along the
LOS). This is too large to be caused by rotation on a galactic scale.
Therefore, we must look to local gas dynamics for an explanation of the
observed motions.  It turns out that an expanding shell of gas gives a
consistent explanation for the observed velocity and column density
pattern (Fig. 2):  closer inspection shows that the components 2 and 3
are weaker in the C spectrum; they are also more widely spaced than in
A,  and they exhibit asymmetric tails pointing towards each other.
Fig. 2 shows (in a schematic drawing) how this pattern can arise if we
are looking along two LOS intersecting an expanding spherical shell,
where one of the LOS intersects the expanding gas more peripherally
than the other.  The velocity shear projected along the LOS is caused
by the projected expansion velocity, $\Delta v_A=2 v_{exp}\cos
\theta_A$ being smaller than $\Delta v_C$ on account of the more
tangential contact of LOS A.  The longer path length of LOS A through
the shell also may account for the higher column density of the
absorption lines in A. The fact that there is a noticeable velocity
difference means that the radius of the shell (always presuming a
spherical homogeneous geometry) is indeed not much larger than  the
transverse beam separation $\Delta r$ = 26 pc.  For a spherical shell,
the radius $R$ is linked to the impact parameters of the LOS from the
center of the shell, $b_A$ and $b_C$, the beam separation $\Delta r$
$(=b_C - b_A)$ and the observed velocity separations between the two
absorption lines supposed to be caused by the expanding shell  (2 and
3)  $\Delta v_A$ and $\Delta v_C$ by
\begin{eqnarray}
R^2 = \frac{(b_C + \Delta r)^2 - b_C \left(\frac{\Delta v_A}{\Delta v_C}\right)^2}
{1 - \left(\frac{\Delta v_A}{\Delta v_C}\right)^2},
\end{eqnarray}
where $\Delta v_A/\Delta v_C$ =0.84 (the velocity separations are measured
between the peaks of the components 2 and 3).
Since $b_C$ can vary between $b_C=0$ and $b_C = R-\Delta r$, we find 

\begin{eqnarray}
13 \leq R \leq 48\ {\mathrm pc}.
\end{eqnarray}

Since the absolut impact parameters of the LOS from the center of the shell
are unknown we can only get a lower
limit to the velocity of expansion $v_{exp}$ from the velocity width measured
between the absorption maxima of components 3 and 4 along LOS C:

\begin{eqnarray}
2 v_{exp} \geq 195\ {\mathrm kms}^{-1}.
\end{eqnarray}

\begin{figure}[tbp]
\centerline{
\psfig{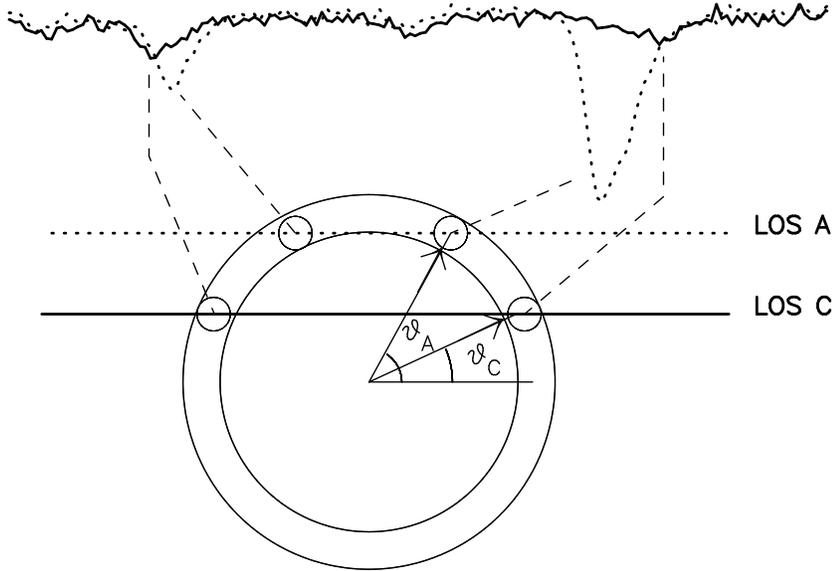}}   
\caption{\small speculative schematic diagram, showing how the low
ionization absorption pattern (CII 1334 region) may arise when two LOS
A (dotted line) and C (solid line) intersect an expanding shell of gas
at different impact parameters and with different angles
$\theta_A$ and $\theta_C$ between the expansion velocity vector and  the LOS. \label{scheme}}
\end{figure}

\section{Discussion}

We can summarize the main findings as follows:  low ionization QSO
absorption systems (as defined by the presence of singly ionized Si, C,
etc) can be very small entities changing significantly over scales of tens
of parsecs. This appears distinct from the properties of the high
ionization gas (CIV systems) the density and velocity structure of
which changes only over separations exceeding a few hundred parsecs
(Rauch 1998, Rauch, Sargent, \& Barlow in prep.). Our directly measured
sizes clearly confirm longstanding predictions based largely on
ionization arguments (Sargent et al.  1979, Bergeron \& Stasinska 1986,
Giroux, Sutherland \& Shull 1994, Petitjean, Riediger \& Rauch 1996)
that the absorption indeed arises in an inhomogeneous gas with several
ionization phases and different spatial extent.

To explain the  finding of velocity differences over tens of parsecs we
must conclude that low ionization QSO absorption systems derive their
internal velocity dispersion from local gas dynamics, either
originating from the host galaxy of the background QSO itself, or from
stellar sources of momentum in an intervening galaxy. While the
relatively high metallicities together with the proximity (in velocity)
to the background QSO are consistent with the gas being associated with
the QSO host galaxy, the low ionization state, full coverage of the QSO
emission regions, and the absence of peculiar abundance patterns argue
against this possibility.

It is hard to say what the precise origin of these absorption features
is.  We have argued here that an expanding gas shell like a bubble
blown by a stellar wind or a supernova, while not being a unique
explanation, is quite consistent with the data. Indeed, Shi (1995) has
drawn attention to the possibility, that supernova type II shells may
explain the large velocity dispersions often seen in CIV systems.
Simulations of the formation of high redshift galaxies indicate that
the velocity width of typical CIV systems is more likely to reflect the
velocity structure of larger, galaxy forming regions (e.g., Rauch,
Haehnelt \& Steinmetz 1997) than individual SN, but we suggest here that a supernova or
stellar wind origin may well be more relevant for the denser
environment of some of the low ionization systems, possibly associated with
galactic disks or the denser regions of halo clouds.  As Shi (1995)
points out, a "standard" type II supernova explosion energy release
(10$^{51}$ ergs) is sufficient to accelerate 10$^4 M_{\odot}$ of matter
to an expansion velocity of 100 kms$^{-1}$, which is well consistent
with the energy requirements in the present case.

Summarizing, it appears that the two-dimensional information
provided by close multiple LOS allows us for the first time to
resolve the spatial and kinematic structure of an individual low
ionization absorption system.  At the same time, this observation
demonstrates that with the new generation of large optical telescopes
we are now in a position to study the interstellar medium at
arbitrarily high redshift. It is likely that such observations will
reveal familiar astrophysical environments like supernova shells or
stellar winds. The much earlier stage in the cosmic development at
z$\sim$ 3.5 (for example, the influence of shallower potential wells on
the gasdynamics, the onset of nucleosynthesis as a function of the
density of the environment, etc) may ultimately show up in larger data
sets of this kind.

Judging from a single case it is not possible to say whether small
scale motions {\em in general} play a role as important as suggested
here, even if most of the low ionization clumps should indeed have a
local, stellar origin.  But if so, the measured velocity dispersion of
an absorption system, consisting of the quadratic sum of the {\em
local} velocity dispersion (for example, $2 v_{exp}$, if caused by an
expanding shell), and the {\em galactic large scale} motion, may
receive a dominant contribution from the former.  In that case attempts
at modelling the kinematics of absorption systems solely in terms of
gravitational motions (e.g., as rotation of galactic disks) may be
overestimating the depth of the underlying galactic potential wells.

Finally, the success of a future statistical study of the high redshift
ISM with this method at present crucially depends on more bright, high
redshift (z$>$ 2) lensed QSOs being found with at least two bright images  and
image separations in excess of 1 -- 1.5 arcsec. Unfortunately, a typical
lensed QSO usually has (if at all) only one image bright enough (V$<$19) for
spectroscopy with a 10m class telescope, and shows image separations of
less than 1.5 arcsec so that the majority of lensed objects are
just not accessible to high resolution observations.  Thus, a promising  
future observational approach, providing the spatial resolution
necessary to split the images and at the same time the needed higher signal-to-noise ratio, will probably
involve adaptive optics in conjunction with an optical or infrared high
resolution spectrograph.

\acknowledgments

We thank the
W.M. Keck Observatory Staff for their help with the observations,
and for their readiness to depart from standard observing procedures,
Alec Boksenberg for correcting the identification of one of the absorption systems, Bob Carswell for useful discussions and for help with the fitting program, and
Martin Haehnelt for reading an earlier draft.
MR is grateful to NASA for support through grant HF-01075.01-94A from
the Space Telescope Science Institute. WLWS and TAB were supported by
grant AST-9529073 from the National Science Foundation. 

\pagebreak

\pagebreak

\begin{deluxetable}{cccccccccccccccc}
\small
\tablewidth{0pt}
\tablenum{1}
\tablecaption{Log. Column Densities of Various Low Ionization Lines}
\tablehead{
\multicolumn{2}{c}{comp.} &
\multicolumn{2}{c}{1} &
\colhead{} &
\multicolumn{2}{c}{2} &
\colhead{} &
\multicolumn{2}{c}{3} &
\\
\tablevspace{.2cm}
\cline{3-4}\cline{6-7}\cline{9-10}\\
\tablevspace{-.2cm}
\colhead{LOS} &
\colhead{} &
\colhead{A} &
\colhead{C} &
\colhead{} &
\colhead{A} &
\colhead{C} &
\colhead{} &
\colhead{A} & 
\colhead{C}\\
\tablevspace{-.4cm}
}
\startdata
\tablevspace{-.1cm}
SiII&&11.53&11.37&&12.13&11.81 &&12.82&11.80\\
&&$\pm$0.11&$\pm$0.08&&$\pm$0.03&$\pm$0.04 &&$\pm$0.01&$\pm$0.05\\  
\\
CII &&12.05 &12.18&&13.14&12.76 &&13.73&12.73  \\
 &&$\pm$0.12 &$\pm$0.14&&$\pm$0.03&$\pm$0.04 &&$\pm$0.01&$\pm$0.06  \\
\\
OI  & &$<$ 12.7 &$<$ 12.7  &&$<$ 12.7 &$<$ 12.7 &&13.27  &$<$13.0  \\
   && &  &&  & &&$\pm0.06$  &  \\
\\
SiIV  & &-- &--  && 12.77 &12.85 &&$<$ 12.0  & $<$ 12.0 \\
   && & && $\pm$0.05 &$\pm$0.06 &&  &  \\
\\
CII/SiII  & &0.52 &0.81  && 1.01 &0.95 &&0.91  &0.93  \\
   &&$\pm$0.16 &$\pm$0.16   && $\pm$0.04  &$\pm$0.06 &&$\pm$0.01  &$\pm$0.08  \\
\enddata
\end{deluxetable}

\end{document}